\documentclass[12pt]{article}

\usepackage{graphicx}

\begin{document}

\title{Reevaluation of the Density Dependence of
Nucleon Radius and Mass in the Global Color Symmetry Model of
QCD\footnote{Work supported by the National Natural Science
Foundation of China, the Major State Basic Research Development
Program and the Foundation for University Key Teacher by the
Ministry of Education}}

\author{{Yu-xin Liu$^{1,2,3,4,5}$, Dong-feng Gao$^{1}$,
Jian-hang Zhou$^{1}$, and Hua Guo$^{2,4}$} \\
\noalign{\vskip 5mm} {\normalsize {$^{1}$ Department of Physics,
Peking University, Beijing 100871, China} }\\
{\normalsize {$^{2}$ The Key Laboratory of Heavy Ion Physics,
Ministry of Education, China, and }} \\
{\normalsize{Department of Technical Physics,
Peking University, Beijing 100871, China} }\\
{\normalsize {$^{3}$ Institute of Theoretical Physics, Academia Sinica,
Beijing 100080, China} }\\
{\normalsize {$^{4}$ Center of Theoretical Nuclear Physics,
National Laboratory of }}\\
{\normalsize {Heavy Ion Accelerator, Lanzhou 730000, China}} \\
{\normalsize{$^5$ CCAST(World Lab.), P. O. Box 8730, Beijing
100080, China}} }

\maketitle


\begin{abstract}
With the global color symmetry model (GCM) at finite chemical
potential, the density dependence of the bag constant, the total
energy and the radius of a nucleon in nuclear matter is
investigated. A relation between the nuclear matter density and
the chemical potential with the action of QCD being taken into
account is obtained. A maximal nuclear matter density for the
existence of the bag with three quarks confined within is given.
The calculated results indicate that, before the maximal density
is reached, the bag constant and the total energy of a nucleon
decrease, and the radius of a nucleon increases slowly, with the
increasing of the nuclear matter density. As the maximal nuclear
matter density is reached, the mass of the nucleon vanishes and
the radius becomes infinite suddenly. It manifests that a phase
transition from nucleons to quarks takes place.

\end{abstract}

\bigskip



\newpage

\parindent=20pt

\section{Introduction}

It is well known that nucleons bound in nuclear medium alter their
properties from those in free space. For instance, an effective
mass ($<M_0$) should be introduced to the nucleon to describe the
property of nuclear matter\cite{SW97}, and a swell hypothesis
should be made to describe the EMC effect well (see for example
Ref.\cite{LSYS94}). Although how much those properties change is
of fundamental interest in nuclear physics, it is still not clear
now. Then, in the relativistic mean field theory (RMF) based on
the $\sigma -\omega$ model, the effective nucleon mass in nuclear
medium is determined with iteration on the fields of nucleon and
mesons\cite{SW97}. However, this model is valid only at the
hadronic level, at which a nucleon is described as a point-like
particle. Because of the importance of the substructure of nucleon
observed in deep inelastic scattering experiments and predicted by
quantum chromodynamics (QCD), it is imperative to employ the quark
and gluon degrees of freedom to describe nuclear phenomena. Since
it is now very difficult to make quantitative predictions with QCD
at low and intermediate energy region, various phenomenological
models based on the QCD assumptions, such as the bag
models\cite{Jaf745,FL778}, quark-meson coupling (QMC)
model\cite{Gui88}, and so on, have been developed. In bag models
and the QMC model, the size and mass of a nucleon are represented
by the radius and energy of the bag with quarks being confined,
respectively. In order to take the modification of the volume
energy on the mass of a nucleon into account, the bag constant is
exploited in the models. To take into account the nuclear matter
medium effect in the QMC model, a phenomenological density
dependence of the bag constant in the medium has been
introduced\cite{LTh98,JJ967,MJ978,Guo99,Su990}. In the global
color symmetry model (GCM)\cite{CR858,FT9127,Tan97,LLZZ98} of QCD,
even though the bag constant, the radius and the mass of a nucleon
in nuclear matter have been evaluated\cite{LGG01} in a consistent
way, a Fermi gas approximation was taken for the relation between
the nuclear matter density and the chemical potential. The
interaction among the nucleons which is believed to be the
residual of the quark-quark interactions has not yet been taken
into account. Then it is still necessary to investigate the
nuclear matter density dependence of the radius, the mass and the
bag constant of a nucleon sophisticatedly.

It has been shown that the global color symmetry model
(GCM)\cite{CR858,FT9127,Tan97,LLZZ98} is a quite successful
effective field theory of QCD in describing hadron properties in
free space(i.e., at temperature $T=0$, chemical potential $\mu
=0$). With the global color symmetry model at finite chemical
potential $\mu$, we will determine the relation between the
nuclear matter density and the chemical potential in the GCM and
reevaluate the nuclear matter density dependence of the bag
constant, the total energy and the radius of a nucleon in this
paper.

The paper is organized as follows. In Section 2 we describe the
formalism of the GCM model at finite chemical potential $\mu$ and
the relation between the chemical potential and the baryon density
in nuclear matter. In Section 3 we represent the calculation and
the obtained results of the bag constant, the bag radius and the
bag energy as functions of the nuclear matter density. In Section
4, a brief summary and some remarks are given.

\section{Formalism}

The starting point of the global color symmetry model (GCM) is the
action in Euclidean matric\cite{CR858}
$$ S=\int d^{4}xd^{4}y\left[ \overline{q}(x)\left( \gamma \cdot
\partial + m \right) \delta(x-y) q(y) - \frac{g^2}{2} j^{a}_{\mu}(x)
 D_{\mu \nu}(x-y) j_{\nu}^{a}(y) \right ] \, , $$
 where $j_{\mu}^{a} = \bar{q}(x) \frac{\lambda^a}{2}\gamma_{\mu} q(x)$ is
the quark current, $D_{\mu \nu}(x-y)$ is an effective two-point
gluon propagator, $m$ is the current quark mass, $g$ is the
quark-gluon coupling constant. Taking the effective gluon
propagator to be diagonal, i.e., $D_{\mu \nu}(x-y) = \delta_{\mu
\nu} D(x-y)$ and applying the transformation of Fierz reordering
to the quark fields, one can rewrite the current-current term as
 $$\frac{g^2}{2} \int d^4 x d^4 y j^{a}_{\mu}(x) D(x-y) j_{\nu}^{a}(y) =
- \frac{g^2}{2} \int d^4x d^4y j^{\theta}(x,y) D(x-y)
j^{\theta}(y,x)\, , $$
 where $j^{\theta}(x,y) = \bar{q}(x) \Lambda^{\theta} q(y)$ with
$\Lambda^{\theta}$ being the direct products of Lorentz, flavor
and color matrices which produce the scalar, vector, pseudoscalar
and axial vector terms. It is obvious that such a $j^{\theta}(x,
y)$ is a bilical current. With two flavors $u$ and $d$ of quarks
being taken into account, each $\Lambda$ is either isoscalar or
isovector. The color matrices involved in the Fierz transformation
contains color-singlet and color-octet terms. Taking the
bosonization procedure one can transfer the bilocal quark current
structure into auxiliary Bose-fields carrying the quantum number
$\theta$. The action of the GCM in free space (i.e., at the
chemical potential $\mu = 0$) for the zero-mass quark can then be
rewritten\cite{CR858} in the Euclidean space as
$$
S(B)=\int d^{4}xd^{4}y\overline{q}(x)\left[\gamma \cdot \partial
\delta(x-y) + \Lambda^{\theta}B^{\theta}(x,y)\right]q(y) + \int
d^{4}xd^{4}y \frac{B^{\theta}(x,y)B^{\theta}(y,x)}{2g^2D(x-y)} \,
, $$
 where $B^{\theta}(x,y)$ is the bilocal Bose-field.
As a consequence, the generating functional is given as
$$Z[\overline{\eta},{\eta}]= {\cal{N}} \int D\overline{q} Dq
DB^{\theta} e^{[-S(B^{\theta}) + \int d^4 x (\overline{\eta}q +
\overline{q}\eta) ] } \, , $$
 where $\overline{\eta}$ and $\eta$ are the quark sources.
To extend the GCM to at finite nuclear matter density (with finite
chemical potential $\mu$), one should, in view of the statistical
mechanics, take the partition function of the canonical (quark)
ensemble into that of the grand ensemble with quarks and hadrons
that are the solitons collecting quarks\cite{FT9127}. The quark
field should be transformed under a constraint on the baryon
number through the chemical potential $\mu$
$$\displaylines{ \hspace*{2cm} q(x) \longrightarrow q^{\prime}(x)
= e^{\mu x_4} q(x) \, . \hfill{(1)} \cr }$$
 After some derivation, we have the action of the GCM in nuclear
matter
$$\displaylines{\hspace*{1cm}  S(B^{\theta}, \mu) = \int \!
d^{4}x d^{4}y \overline{q}^{\prime}(x)\!\left[(\gamma \! \cdot \!
\partial \! - \! \mu\gamma_{4})\delta(x\! - \! y) \! + \!
e^{\mu x_4} \Lambda^{\theta} B^{\theta}(x,y) e^{ -\mu y_4} \right]
q^{\prime} (y) \hfill \cr \hspace*{35mm}  +  \int \! d^{4}xd^{4}y
\frac{B^{\theta}(x,y) B^{\theta}(y,x)} {2g^2D(x-y)} \, , \hfill{}
\cr }$$
 and the generating functional is given as
$$\displaylines{\hspace*{2cm}
Z[\mu , \overline{\eta},{\eta}]=\int D\overline{q} Dq DB^{\theta}
e^{[- S(B^{\theta},\ \mu) + \int \! d^4 x ( \overline{\eta}q +
\overline{q}\eta) ] } \, . \hfill{(2)} \cr }
$$
 After integrating the quark fields, we obtain
$$\displaylines{\hspace*{2mm}
S(B^{\theta}, \mu) \! = \! - \mbox{Tr}{\ln\left[( \gamma\! \cdot
\! \partial \! - \! \mu\gamma_{4}) \delta(x\! - \! y) \! + \!
e^{\mu x_4} \Lambda^{\theta} B^{\theta} e^{-\mu y_4}\right]} \! +
\! \int d^{4}xd^{4}y \frac{B^{\theta}(x,y)B^{\theta}(y,x)}
{2g^2D(x-y)} \, . \hfill{(3)} \cr }$$
 Generally, the bilocal field $B^{\theta}(x,y)$ can be written
as\cite{LLZZ98}
$$\displaylines{\hspace*{20mm}
B^{\theta}(x,y) = B^{\theta}_0(x,y) + \sum_{i} \Gamma^{\theta}_{i}
(x,y) \phi^{\theta}_{i}(\frac{x+y}{2}) \, , \hfill{(4)} \cr }$$
 where $B^{\theta}_{0}(x,y)\! = \! B^{\theta}_0(x-y) = B^{\theta}(x-y)$
is the vacuum configuration of the bilocal field. ${\Sigma_{i}
\Gamma^{\theta}_{i}(x,y) \phi^{\theta}_{i}(\frac{x + y}{2}) }$
correspond to the fields which can be interpreted as effective
meson fields. In the lowest order approximation with only the
Goldstone bosons being taken into account, the $\phi^{\theta}_{i}$
includes $\pi$ and $\sigma$ mesons. The corresponding width of the
fluctuations are approximately the same as the vacuum
configuration\cite{LLZZ98}, i.e., $\Gamma^{\theta}_{i} =
B^{\theta}_{0}$.  Since the bilocal field arises from the bilocal
current of quarks, the internal $\overline{q}$-$q$ structure of
the mesons can be well described in the GCM. The vacuum
configuration can be determined by the saddle-point condition
$\frac{\partial S}{\partial B^{\theta}_0}=0$. Then an equation of
the translational invariant quark self-energy $\Sigma (q,\mu )$ is
obtained as a truncated Dyson-Schwinger equation
$$\displaylines{\hspace*{15mm}
\Sigma(p,\mu)=\int \frac{d^{4}q}{(2\pi)^4}g^2D(p-q)\frac{t^{a}}{2}
\gamma_{\nu} \frac{1}{i\gamma \cdot q-\mu\gamma_{4}+\Sigma(q,\mu)}
\gamma_{\nu} \frac{t^{a}}{2}\, , \hfill{} \cr }$$
 With $\tilde{q}_{\mu}=(\vec{q},q_4 + i\mu)$ being introduced, the
above equation can be rewritten as
$$\displaylines{\hspace*{15mm}
\Sigma(\tilde{p})=\int
\frac{d^{4}q}{(2\pi)^4}g^2D(p-q)\frac{t^{a}}{2}
\gamma_{\nu}\frac{1}{i\gamma \cdot \tilde{q}+\Sigma(\tilde{q} )}
\gamma_{\nu} \frac{t^{a}}{2}\, . \hfill{(5)} \cr }$$
 Considering the fact that the inclusion of the chemical potential
breaks the O(4) symmetry in the four-dimensioanl space, one should
rewrite the decomposition of the self-energy $\Sigma$  as
$$\displaylines{\hspace*{20mm}
\Sigma(p,\mu)=i[A(p,\mu)-1]\vec{\gamma} \cdot \vec{p} + i[C(p,\mu)
-1]\gamma_4 (p_4 + i \mu) + B(p,\mu)\, , \hfill{(6)} \cr }
$$
 where $B(p, \mu)$ is the counterpart of the vacuum configuration
in the momentum space, i.e.,
 $$ B(p, \mu) = \int \frac{1}{ (2 \pi)^4} B^{\theta}_{0} (z) e
^{i \tilde{p} z} d z \, .$$
 It requires then that the vacuum configuration of $\sigma$ and $\vec{\pi}$
should satisfy the restriction $\sigma ^2 + \vec{\pi} ^2 = 1 $.
 Combining Eqs.~(5) and (6), one can obtain the equations to
determine the $A(\tilde{p})$, $B(\tilde{p})$ and $C(\tilde{p})$ as
follows
$$\displaylines{\hspace*{15mm}
\left[A(\tilde{p})-1\right]\vec{p}^2 = \frac{8}{3}\int
\frac{d^{4}q}{(2\pi)^4} g^2 D(p-q)\frac{A(\tilde{q})\vec{q}\cdot
\vec{p}} {A^2(\tilde{q}) \vec{q}^2 + C^2(\tilde{q}) \tilde{q}_4^2
+ B^2(\tilde{q})} \, , \hfill{(7a)} \cr \hspace*{15mm}
\left[C(\tilde{p})-1\right]\tilde{p}_4^2 = \frac{8}{3}\int
\frac{d^{4}q}{(2\pi)^4} g^2 D(p-q)\frac{C(\tilde{q})\tilde{q}_4
\tilde{p}_4} {A^2(\tilde{q}) \vec{q}^2 + C^2(\tilde{q})
\tilde{q}_4^2 + B^2(\tilde{q})} \, , \hfill{(7b)} \cr
\hspace*{29mm}
 B(\tilde{p})=\frac{16}{3}\int\frac{d^{4}q}{(2\pi)^4}g^2D(p-q)
 \frac{B(\tilde{q})}{A^2(\tilde{q})\vec{q}^2 + C^2(\tilde{q})
 \tilde{q}^2_4 + B^2(\tilde{q})} \, . \hfill{(7c)} \cr } $$

Basing on the solution of the Dyson-Schwinger equations, one can
determine the bilocal field, and fix further the GCM action.

With a nontopological-soliton ansatz, the action of the bag models
at finite chemical potential $\mu$ in terms of the GCM can be
given by extending the formalism proposed in Refs.\cite{CR858} and
\cite{FT9127} as
$$\displaylines{\hspace*{1cm}
S_B=\overline{q}_{j}\{\gamma \cdot
\partial- \gamma_4 \mu - \alpha [\sigma(x)-i\vec{\pi}(x) \cdot
\vec{\tau}\gamma_5 ] \} q_{j} +\hat{S}(\sigma,\pi,\mu) \, ,
\hspace*{1cm} (j=1,2,3), \hfill{(8)} \cr }$$
 where $\hat{S}(\sigma,\pi,\mu)$ includes $\sigma$ and $\pi$ mesons
and reads
$$\displaylines{\hspace*{8mm}
\hat{S}(\sigma,\pi,\mu)\! = \! \int[\frac{1}{2} m_{\sigma} ^2
\sigma ^2 + \frac{f_{\sigma}^2}{2} (\partial_{\mu}\sigma)^2 +
\frac{1}{2} m_{\pi}^2 \pi^2 + \frac{f_{\pi}^2}{2} (\partial_{\mu}
\vec{\pi})^2 \! - \! V(\sigma,\pi)]d^{4}z +\cdots \hfill{(9)} \cr
}$$
 with
$$\displaylines{\hspace*{10mm}
V(\sigma,\pi)=- 12 \int \frac{d^{4} p}{(2\pi)^4} \left\{ \ln
\left[\frac{A^{2}(\tilde{p}) \vec{p}^2+ C^{2}(\tilde{p})
\tilde{p}_4^2 +
(\sigma^2+\vec{\pi}^2)B^{2}(\tilde{p})}{A^{2}(\tilde{p}) \vec{p}^2
+ C^{2}(\tilde{p}) \tilde{p}_4^2 + B^{2}(\tilde{p})} \right]
\right. \hfill{} \cr \hspace*{20mm} \left.
-\frac{(\sigma^2+\vec{\pi}^2-1) B^{2}(\tilde{p})}{A^{2}(\tilde{p})
\vec{p}^2 + C^{2}(\tilde{p}) \tilde{p}_4^2 +B^{2}(\tilde{p})}
\right\}\, , \cr } $$
 and the quark meson coupling constant $\alpha$ is given as
$$ \alpha(x) = \int \frac{d^{4} p}{(2 \pi)^4} B(\tilde{p}) e^{-i
\tilde{p} \cdot x} \, .$$
 It is evident that such a quark meson coupling constant is just
the vacuum configuration of the bilocal field and is independent
of the meson fields.

Analyzing the stationary property of the bag and differentiating
Eq.~(8), one has the equations of motion for the quarks and mesons
$$\displaylines{\hspace*{2cm} \left\{ \gamma \cdot \partial  -
\gamma_4 \mu - \alpha [ \sigma(x) - i \vec{\pi}(x) \cdot
\vec{\tau}\gamma_5 ] \right\} q_{j} = 0 \, , \hfill{(10)} \cr }$$
$$\displaylines{\hspace*{3cm} \frac{\partial S_B}{\partial \sigma(x)} = 0
\, , \hfill{(11)} \cr }$$
 $$\displaylines{\hspace*{3cm} \frac{\partial
S_B}{\partial \pi(x)} = 0\, . \hfill{(12)} \cr }$$
 The quark field and $\sigma$, $\pi $ meson fields in symmetric nuclear
matter can be determined by solving the Eqs.~(10-12)
self-consistently. As a consequence, the corresponding energies
can be obtained. It is apparent that the meson fields
corresponding to the vacuum configuration can be simply taken as
$\sigma =1$, $\pi = 0$ due to the restriction $\pi ^2 + \sigma ^2
=1$. In light of the nontopological-soliton
ansatz\cite{FL778,CR858}, one can approximately take the meson
fields inside a bag (i.e., in a nucleon) as $\sigma=0 $ and
$\pi=0$. For the quarks in a bag, Eq.~(10) can thus be rewritten
approximately as
 $$\displaylines{\hspace*{3cm} \left[
\gamma \cdot \partial - \gamma_4 \mu \right] q_j(x) = 0 \, .
\hfill{(13)} \cr }$$
 The lowest total energy of a single quark
with respect to the radius $R$ of the bag is given as
$$\displaylines{\hspace*{3cm} \epsilon_{j}(R)=\frac{\omega_0 - \mu }{R} ,
\hfill{(14)} \cr }$$
 where $\omega_0=2.04 .$  For a free nucleon (i.e., with $\mu =0$),
this result is consistent with those obtained in
Refs\cite{Jaf745,LTh98,JJ967}. And the bag constant ${\cal{B}}$ is
obtained as
$$\displaylines{\hspace*{5mm}
{\cal{B}}=V(0,0,) - V(1,0) = 12 \int \!\!\frac{d^4 p}{(2\pi)^4}
\left\{\ln \left[ \frac{A^{2}(\tilde{p})\vec{p}^2\! + \!
C^{2}(\tilde{p})\tilde{p}_{4}^2\! + \! B^{2}(\tilde{p})}
{A^{2}(\tilde{p})\vec{p}^2 \! + \! C^{2}(\tilde{p})\tilde{p}_{4}^2
}\right] \right. \hfill{} \cr \hspace*{6cm}  \left. -
\frac{B^{2}(\tilde{p})} {A^{2}(\tilde{p})\vec{p}^2 \! + \!
C^{2}(\tilde{p})\tilde{p}_{4}^2 \! + \! B^{2}(\tilde{p})} \right\}
\, . \hfill{(15)} \cr }$$
 With the correction from the motion of center-of-mass, the zero-point
effect and the color-electronic and color-magnetic interactions
being taken into account, the total energy of a bag is given as
$$\displaylines{\hspace*{2cm}
E=3\epsilon_{j}(R)+\frac{4}{3}\pi R^{3}{\cal{B}}-\frac{Z_0}{R}
 = \frac{3(\omega_0 - \mu) - Z_0}{R}+\frac{4}{3}\pi R^{3}{\cal{B}} \, ,
\hfill{(16)} \cr }$$
 where $ Z_0/R$ denotes the corrections of the
motion of center-of-mass, zero-point energy and other effects.

Just as the same as that in Ref.\cite{CR858}, the bag is
identified as a nucleon in the present work. It satisfies the
equilibrium condition
$$ \frac{dE(R)}{dR} = 0 \, . $$
From this condition, we get
$$\displaylines{\hspace*{3cm}
R=\left(\frac{a}{4\pi {\cal{B}}}\right)^{1/4} \, , \hfill{(17)}
\cr }$$
 where $ a=3(\omega_0 - \mu)-Z_0 . $ As a consequence, Eq.~(16) can
be rewritten as
 $$\displaylines{\hspace*{3cm} E=\frac{4a}{3} \left(
\frac{4\pi {\cal{B}}}{a} \right)^{1/4} \, . \hfill{(18)} \cr }$$

It is apparent that, with the solutions of Dyson-Schwinger
equations (Eqs.~(7a), (7b) and (7c)) being taken as the input for
Eqs.~(15), (17) and (18), the properties of nucleons (i.e., bags)
in nuclear matter can be obtained.

In the practical calculation, since the knowledge about the exact
behavior of $ g^2$ and $D(p-q)$ in low energy region is still lacking,
one has to take some approximations or phenomenological form to solve
the Dyson-Schwinger equations. For simplicity, we adopt the infrared
dominative form\cite{MN83,CR858}
 $$\displaylines{\hspace*{4cm}
g^2D(p-q)=\frac{3}{16}\eta^2 \delta(p-q) \, , \hfill{(19)} \cr }$$
 where $\eta$ is an energy-scale parameter and can be fixed by
experimental data of hadrons. Although this form does not include
the contribution from the ultraviolet energy region, it maintains
the main property of QCD in the low energy region. With Eqs.~(7a),
(7b), (7c) and (19), one has
$$\displaylines{\hspace*{5mm}
A(\tilde{p})=C(\tilde{p})=2, \qquad \qquad
B(\tilde{p})=(\eta^2-4\tilde{p}^2)^{1/2}, \qquad \qquad \mbox{for}
\; Re(\tilde{p}^2)< \frac{\eta^2}{4}, \hfill{(20a)} \cr
\hspace*{5mm}
A(\tilde{p})=C(\tilde{p})=\frac{1}{2}\!\left[1\!+\!\left( 1\!+\!
\frac{2\eta^2}{\tilde{p}^2} \right)^{1/2} \right],  \ \
B(\tilde{p})=0, \qquad \qquad \mbox{for} \; Re(\tilde{p}^2)
> \frac{\eta^2}{4}, \hfill{(20b)} \cr }  $$
 which describes the phase where chiral symmetry is spontaneously
broken and the dressed-quarks are
confined\cite{FT9127,BR96,BRS98,MRS98,BGP99}. Meanwhile one has
also the Wigner solution
$$\displaylines{\hspace*{10mm}
A(\tilde{p})=C(\tilde{p})=\frac{1}{2}\!\left[1\!+\!\left( 1\!+\!
\frac{2\eta^2}{\tilde{p}^2} \right)^{1/2} \right],  \ \
B(\tilde{p})=0,  \hfill{(21)} \cr }  $$
 which characterizes a phase in which chiral symmetry is not
broken and the dressed-quarks are not confined.

In order to investigate the dependence of nucleon properties on
the nuclear matter density $\rho$ explicitly, we must transfer the
above obtained $\mu$-dependence to that of the $\rho$-dependence.
It has been known that the baryon number in nuclear matter can be
related with the generating functional $Z(B^{\theta}, \mu)$ of the
system\cite{FT9127,Kap89,Lu02} as
 $$\displaylines{\hspace*{2cm}
 n = \frac{\partial}{\partial \mu} \ln Z(B^{\theta}, \mu ) \, .
 \hfill{(22)} \cr }$$
 Combining Eqs.~(2), (3) and (8), and accomplishing a Legendre
transformation along the way described in Ref.\cite{FT9127}, we
have
$$\displaylines{\hspace*{3cm}
n \approx - \frac{\partial}{\partial \mu} S( B^{\theta}, \mu ) =
\frac{\partial}{\partial \mu} Tr \ln G^{-1} ( B^{\theta}, \mu )\,
, \hfill{(23)} \cr }$$
 where $G^{-1}(B^{\theta}, \mu )$ is the inverse of the quark
propagator in the medium, i.e.,
 $$\displaylines{\hspace*{2cm}
G^{-1}(B^{\theta}, \mu ) = ( \rlap/{\partial} - \gamma_4 \mu )
\delta(x-y) + e^{\mu x_4} \Lambda ^{\theta}B^{\theta}(x-y) e^{-
\mu y_4} \, . \hfill{(24)} \cr }$$
 Extending $\ln G^{-1}(B^{\theta}(x-y), \mu )$ in a Taylor series,
we have
$$\displaylines{\hspace*{5mm}
\ln G^{-1}(B^{\theta}(x-y), \mu ) = \ln[(\rlap/{\partial} - \gamma
_4 \mu )\delta(x-y) + e^{\mu x_4} \Lambda ^{\theta}B^{\theta}(x-y)
e^{-\mu y_4} ] \hfill{} \cr \hspace*{4.3cm} =
\ln[(\rlap/{\partial} - \gamma _4 \mu ) \delta(x-y)] + \Big\{
\frac{1}{\rlap/{\partial} - \gamma _4 \mu} \Lambda
^{\theta}B^{\theta}(x-y) e^{\mu (x_4 - y_4)} \hfill{} \cr
\hspace*{5.5cm} - \frac{1}{2} \Big[ \frac{1} {\rlap/{\partial} -
\gamma _4 \mu} \Lambda ^{\theta}B^{\theta} (x-y) e^{\mu (x_4 -
y_4)} \Big]^{2} + \cdots \Big\} \, . \hfill{} \cr }$$
 We get then
$$\displaylines{\hspace*{1cm}
n \approx - \frac{1}{2} \frac{\partial}{\partial \mu} tr \int
d^{4}y d^{4}x \frac{1} {\rlap/{\partial} - \gamma _4 \mu} \Lambda
^{\theta}B^{\theta}(x-y) e^{\mu (x_4 - y_4)}  \frac{1}
{\rlap/{\partial} - \gamma _4 \mu} \Lambda
^{\theta}B^{\theta}(y-x) e^{\mu (y_4 - x_4)}\, . \hfill{} \cr }$$
 Considering the Fourier transformation
$$\displaylines{\hspace*{1cm}
\frac{1} {\rlap/{\partial} - \gamma _4 \mu} \Lambda
^{\theta}B^{\theta}(x-y) e^{\mu (x_4 - y_4)} = \int
\frac{d^{4}q}{(2 \pi)^{4}} \frac{\Lambda^{\theta} } {i \rlap/{q}}
B^{\theta}(\vec{q}, q_4 + i \mu) e^{i q (x - y)}\, , \hfill{} \cr
}$$
 we obtain finally
$$\displaylines{\hspace*{1cm}
n \approx 2 \frac{\partial}{\partial \mu} \int \frac{d^{4}x}{(2
\pi)^{4}} \int d^{4}p \frac{[B^{\theta} (\vec{p}, p_4 + i \mu)]
^{2} }{p^2} \, . \hfill{} \cr }$$
 The baryon number density can thus be given as
$$\displaylines{\hspace*{1cm}
\rho_{n} = \frac{n}{\int d^{4} x }\approx \frac{2}{(2 \pi)^4}
\frac{\partial}{\partial \mu} \int d^{4}p \frac{[B^{\theta}(
\vec{p}, p_4 + i \mu)]^{2} }{p^2} \, . \hfill{(25)} \cr }$$

Combining Eqs.~(15), (17), (18), (20) and (25), we can obtain the
dependence of the bag constant ${\cal{B}}$, the total energy $E$
and the radius $R$ of a nucleon (i.e., those of a bag) on the
nuclear matter density $ \rho_{n}$.

\section{Calculation and Results}

By calibrating the nucleon mass $M_0=939$~MeV as the total energy
of a bag and radius $R_0 =0.8$~fm in free sapce (i.e., $\mu =0$,
$\rho =0$), we get the energy-scale $\eta=1.220$~GeV, $Z_0=3.303$.
Such a best fitted energy-scale $\eta$ fits well the value
$1.37$~GeV, which was fixed by a good description of $\pi$ and
$\rho$ meson masses\cite{BRS98,MRS98}, and is much more close to
the Bjorken-scale $1.0$~GeV (see Ref.\cite{CP75} and the
references therein). The obtained $ Z_0$ is larger than the
originally fitted value $1.84$\cite{HK78}. However what we refer
to here includes all the effects but not only the zero-point
energy. Meanwhile other investigations (see for example
Ref.\cite{MK01}) have shown that the zero-point energy parameter
can be larger than 1.84, even though the other effects are taken
into account separately. With the above parameters $\eta$, $Z_0$
and Eq.(15), we get at first the bag constant in free space as $
{\cal{B}}_0 =(172~\mbox{MeV})^4$. It is evident that the presently
obtained value ${\cal{B}}_0$ is quite close to the result given in
Ref.\cite{JJ967}.

By varying the chemical potential $\mu$, we obtain the relation
between the nuclear matter density and the chemical potential from
Eqs.~(20) and (25)
$$ \rho_{n} = \frac{1}{\pi^2} \Big[ \mu ^3 + \frac{\eta ^2}{4}
\mu \Big] \, .$$
 It is obvious that the nuclear matter density changes monotonously
with respect to the chemical potential $\mu$. Such a result shows
also that, besides the global coefficient, the chemical potential
dependence of the baryon density in the GCM differs from that in
the Fermi gas approximation ($\rho_n =\frac{2}{3 \pi^2} \mu^3 $ )
in a linear term which arises from the self-interaction of quarks
(the constituents of nucleons). In the case of small chemical
potential (or low density), this difference can be quite large
since the $\frac{\eta^2}{4}$ may be much larger than $\mu ^2$.
Furthermore, we get the variation behavior of the ratios of bag
constant, the nucleon radius and the total energy of the bag in
nuclear matter to the corresponding value in free space against
the nuclear matter density. The results are illustrated in
Figs.~1-3, respectively. To show the influence of the quark
self-interaction, we show also the results without the interaction
being taken into account (i.e., in the Fermi gas approximation) in
the figures.

From the figures, one may easily realize that, as the density of
nuclear matter increases, the bag constant and the total energy of
the bag (i.e., the mass of a nucleon) decrease monotonously.
Meanwhile, the radius of a nucleon increases. When the nuclear
matter density reaches the value about 12 times the normal nuclear
matter density $\rho _0$ (corresponding to a chemical potential
$\mu \approx 0.316$~GeV), the bag constant and the total energy of
a bag vanish simultaneously, and the radius of a nucleon becomes
infinite. Such behaviors indicate that nucleons can no longer
exist as bags consisting of quarks, i.e., a phase transition from
hadrons to quarks happens. It manifests that a critical nuclear
matter density $\rho_{c} \approx 12 \rho_{0}$ exists for the quark
deconfinement to take place. Since the density $12\rho_0$ is
approximately close to the maximal density in the center area of
neutron stars and larger than the average(which is commonly
believed to be about $10\rho_0$), the presently obtained results
show that the nuclear matter would change to quark matter as the
density of nuclear matter gets beyond the maximal average density
of neutron stars. It provides then a clue that there may be hybrid
matter with both hadrons and quarks, even pure quark matter in the
center part of neutron stars. Such a feature of neutron star is
quite similar to the previous result (see for example
Ref.\cite{Pra97}). Furthermore, pure bare quark stars can also
exist if the matter density is very large (at least $> 12
\rho_{0}$).  It is also worth mentioning that the increase of the
nucleon radius induces naturally a swell of nucleons. The nucleon
swell hypothesis to the EMC effect (see for example
Ref.\cite{LSYS94}) has then a quite solid QCD foundation.

Comparing the results with and without the self-interaction among
quarks, one can easily recognize that the deconfinement phase
transition from hadrons to quark-gluons can happen in the nuclear
matter with a density only little larger than the normal density
if the self-interaction has not been taken into account (i.e., in
the Fermi gas approximation). It indicates that the
self-interaction plays vital role for hadrons to exist as bags
collecting quarks in nuclear matter with rather high density.

Looking over the figures 1-3 more cautiously, one may recognize
that, as the nuclear matter density is much smaller than the
critical density $\rho_{c}$, the radius of the bag is almost
independent of the variation of the nuclear matter density, and
the total energy does not change drastically with respect to the
density either. As the nuclear matter density is close to the
critical value, the total energy and the radius change against the
density abruptly. Furthermore the total energy becomes zero and
the radius gets to be infinite suddenly as the density reaches the
critical value. It shows that the phase transition from hadron
matter to quark matter is the first order phase transition.

It is also worthy mentioning that the presently obtained changing
feature of the radius, the mass and the bag constant of a nucleon
with respect to the nuclear matter density is qualitatively
consistent with that given in the QMC model\cite{LTh98,JJ967,
MJ978,Guo99}. Such a behavior indicates that, dealing the bag
constant and the bag radius with a phenomenological dependence on
the medium density in the QMC model is reasonable.

\section{Summary and Remarks}

In summary we have investigated the density dependence of the bag
constant of nucleons, the nucleon radius and the total energy of
the bag in nuclear matter in the global color symmetry model, an
effective field theory model of QCD. A relation between the
nuclear matter density and chemical potential is fixed with the
GCM action. A maximal density of nuclear matter, which is about 12
times the normal nuclear matter density, for the nucleon to exist
as bags consisting of quarks is obtained. The calculated results
indicate that the bag constant and the total energy of the bag
decreases with the increasing of the nuclear matter density before
the maximal density is reached. Meanwhile the size of nucleons
swells. As the maximal density is reached, a phase transition from
nucleons to quark-gluons takes place. On the other hand, the
calculated results show that the self-interaction among quarks
plays dominant role for baryons to exist as bags consisting of
quarks in the nuclear matter at rather high density. Furthermore,
the presently obtained changing features agree quite well with
those obtained in QMC. In this sense, it provides a clue of the
QCD foundation to the QMC with a simple effective field theory
model of QCD.

In the present calculation, the $g^2D(p-q)$ is taken to be
proportional to a $\delta-$function. However, the detailed effects
of the running coupling constant, the gluon propagator $D(p-q)$
and the other degrees of freedom on the changing feature have not
yet been included. Especially, since the meson fields were taken
to be $\sigma = 0$ and $\pi = 0$ in the bag with respect to those
of the vacuum configuration $\sigma =1$, $\pi =0$, the
self-consistent interaction and adjustment between the quarks and
the meson fields have not yet been taken into account.  Then a
more sophisticated investigation is necessary and under progress.

\bigskip

This work is supported by the National Natural Science Foundation
of China under contact No. 10075002, 10135030 and the Major State
Basic Research Development Program under contract No.G2000077400.
One of the author (Liu) thanks also the support of the Foundation
for University Key Teacher by the Ministry of Education, China.
The authors are also indebted to Professors Fan Wang, Ru-keng Su
and Xiao-fu L\"{u} for their helpful discussions.

\newpage

{\large \bf Figures and Figure  Captions:}

\vspace{15mm}

\begin{center}
\begin{figure}[h]
\includegraphics[scale=0.75,angle=0]{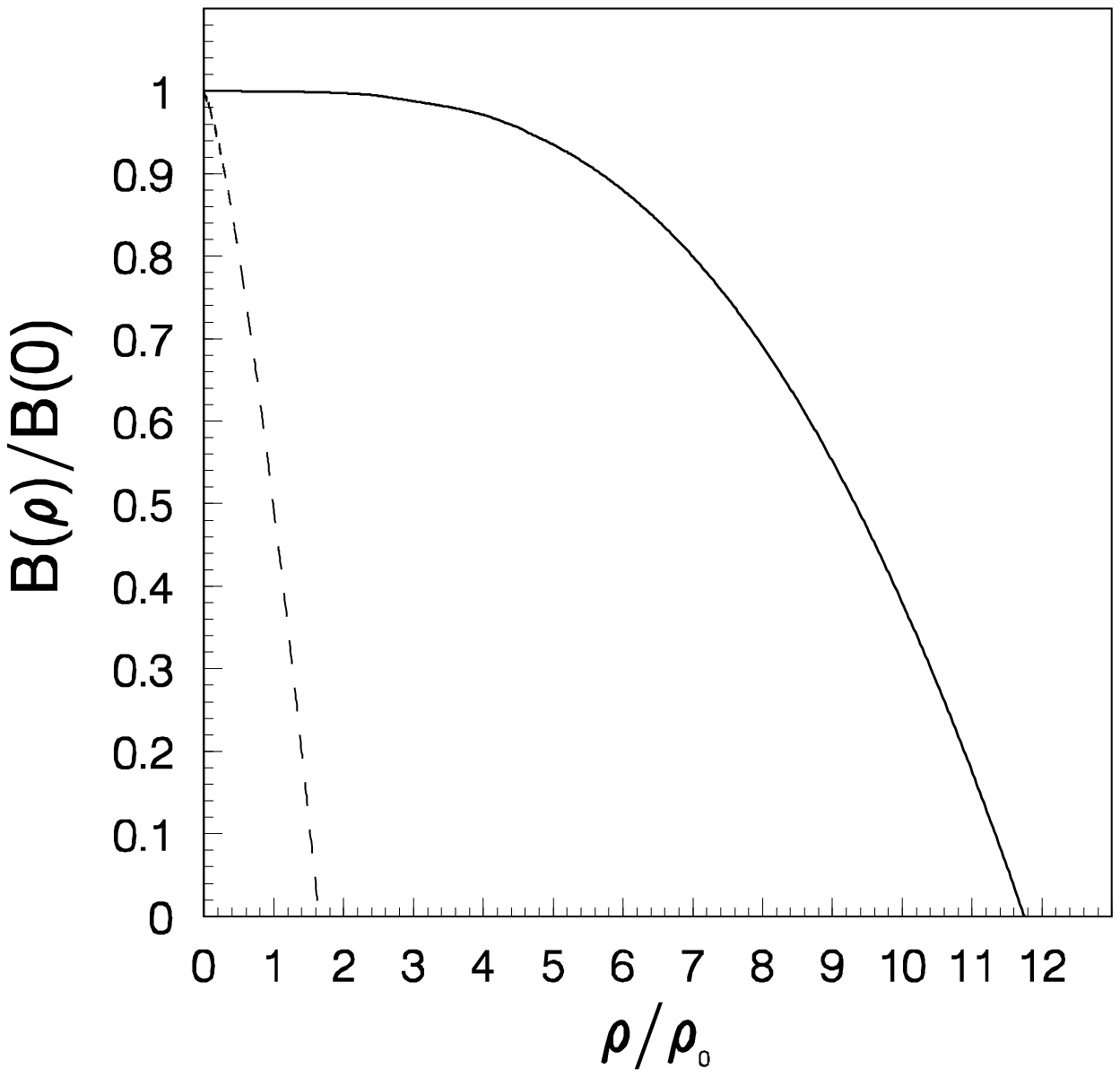}
\caption{Calculated ratio of the bag constant in nuclear matter to
that in free space as a function of the nuclear matter density
(solid curve). The dashed curve illustrates the ratio in Fermi gas
approximation.}
\end{figure}
\end{center}

\begin{center}
\begin{figure}
\includegraphics[scale=0.8,angle=0]{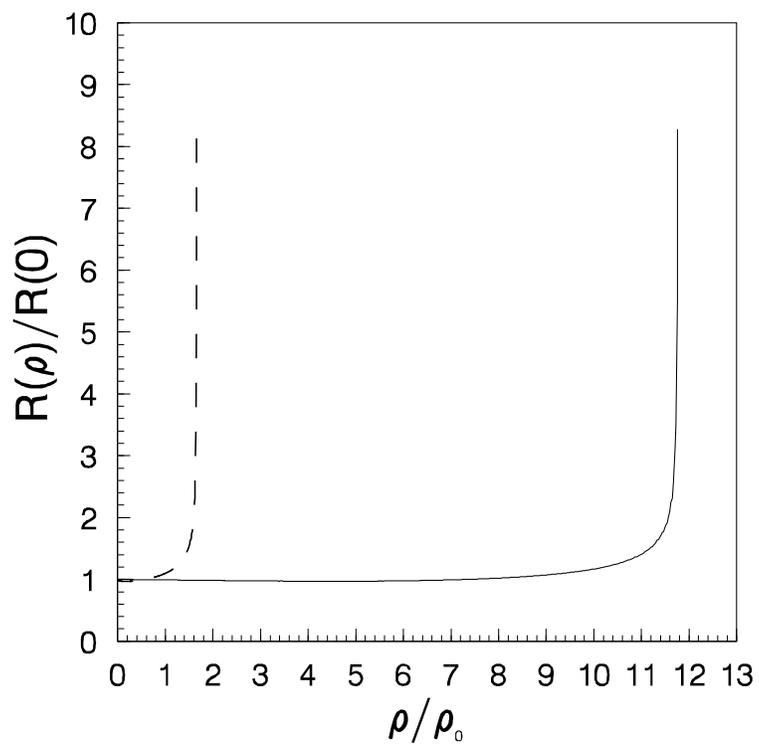}
\caption{Calculated ratio of the radius of nucleon in nuclear
matter to that in free space as a function of the nuclear matter
density (solid curve). The dashed curve represents the ratio in
Fermi gas approximation.}
\end{figure}
\end{center}

\begin{center}
\begin{figure}
\includegraphics[scale=0.8,angle=0]{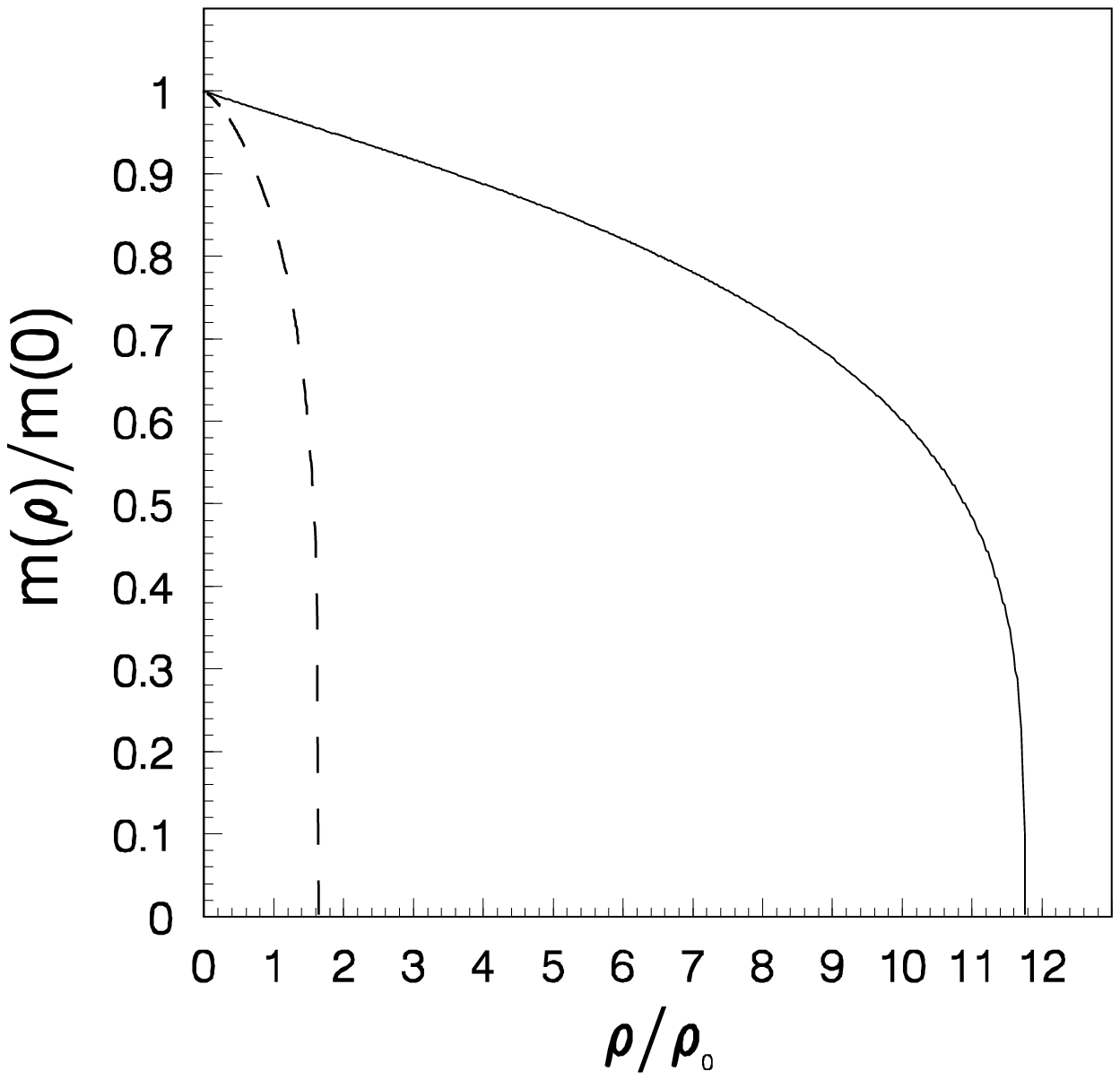}
\caption{Calculated ratio of the total energy of a bag in nuclear
matter to that in free space as a function of the nuclear matter
density (solid curve). The dashed curve displays the ratio in
Fermi gas approximation. }
\end{figure}
\end{center}



\newpage

\begin{thebibliography}{50}
\bibitem{SW97} B. D. Serot and J. D. Walacka, Int. J. Mod. Phys. {\bf E
6} (1997) 515.  \vspace*{-1mm}
\bibitem{LSYS94} G.L. Li, J.P Shen, J.J. Yang and H.Q. Shen, Phys. Rep.
{\bf 242} (1994) 505.
\bibitem{Jaf745} A.Chodos,R. L. Jaffe, K. Jonson, C. B. Thern and
V. F. Weisskopf, Phys. Rev. {\bf D 9} (1974) 3471; A. Chodos, R.L.
Jaffe, K. Jonson and C. B. Thern, Phys. Rev. {\bf D 10} (1974)
2599; T. A. DeGrand, R. L. Jaffe, K. Jonson and J. Kiskis, Phys.
Rev. {\bf D 12} (1975) 2060.
\bibitem{FL778} R. Friedberg and T. D. Lee, Phys. Rev. {\bf D 15} (1977)
1694; {\it ibid}, {\bf D 16} (1977) 1096; {\it ibid}, {\bf D 18}
(1978) 2623.
\bibitem{Gui88} P. A. Guichon, Phys. Lett. {\bf B 200} (1988) 235.
\bibitem{LTh98} D. H. Lu, K. Tsushima, A. W. Thomas, A. G. Williams and
K. Saito, Nucl. Phys. {\bf A 634} (1998) 443.
\bibitem{JJ967} X. Jin and B. K. Jennings, Phys. Lett. {\bf B 374}
(1996) 13; X. Jin and B. K. Jennings, Phys. Rev. {\bf C 54} (1996)
1427;  {\it ibid}, {\bf C 55} (1997) 1567.
\bibitem{MJ978} H. M\"{u}ller, and B. K. Jennings, Nucl. Phys. {\bf A
626} (1997), 966; {\it ibid}, {\bf A 640} (1998) 55.
\bibitem{Guo99} H. Guo, J. Phys. {\bf G 25} (1999) 1701.
\bibitem{Su990} P. Wang, R. K. Su, H. Q. Song and L. L. Zhang, Nucl. Phys.
{\bf A 653} (1999) 166; L. L. Zhang, H. Q. Song, P. Wang and R. K.
Su, J. Phys. {\bf G 26} (2000) 1301.
\bibitem{CR858} R. T. Cahill , C. D. Roberts, Phys. Rev. {\bf D
32} (1985) 2419;  J. Praschifka, C. D. Roberts, and R. T. Cahill,
Phys. Rev. {\bf D 36} (1987) 209; R. T. Cahill, C. D. Roberts, and
J. Praschifka, Ann. Phys. (N.Y.) {\bf 188} (1988) 20; C. D.
Roberts, R. T. Cahill, M. E. Sevior and N. Iannella, Phys. Rev.
{\bf D 49} (1994) 125.
\bibitem{FT9127} M. R. Frank, P. C. Tandy, and G. Fai, Phys. Rev.
{\bf C 43} (1991) 2808; M. R. Frank and P. C. Tandy, Phys. Rev.
{\bf C 46} (1992) 338; C. W. Johnson and G. Fai, Phys. Rev. {\bf C
56} (1997) 3353.
\bibitem{Tan97} P. C. Tandy, Prog. Part. Nucl. Phys. {\bf 39}
(1997) 177.
\bibitem{LLZZ98} X. F. L\"{u}, Y. X. Liu, H. S. Zong and E. G. Zhao,
Phys. Rev. {\bf C 58} (1998) 1195.
\bibitem{LGG01} Y. X. Liu, D. F. Gao, and H. Guo, Nucl. Phys. {\bf
A 695} (2001) 353.
\bibitem{MN83} H. J. Munczek, A. M. Nemirovsky, Phys. Rev. {\bf D
28} (1983) 181.
\bibitem{BR96} A. Bender, D. Blaschke, Y. Kalinovsky, and C. D. Roberts,
Phys. Rev. Lett. {\bf 77} (1996) 3724; A. Bender, G. I. Poulies,
C. D. Roberts, S. Schmidt, and A. W. Thomas, Phys. Lett. {\bf B
431} (1998) 263.
\bibitem{BRS98} D. Blaschke, C. D. Roberts, S. Schmidt, Phys. Lett.
{\bf B 425} (1998) 232.
\bibitem{MRS98} P. Maris, C. D. Roberts, S. Schmidt, Phys. Rev. {\bf C
57} (1998) 2821.
\bibitem{BGP99} D. Blaschke, H. Grigorian, G. Poghosyan, Phys. Lett.
{\bf B 450} (1999) 207.
\bibitem{Kap89} J. I. Kapusta, {\it Finite-Tempetature Field Theory}
(Cambridge University Press, 1989)
\bibitem{Lu02} L.F. Yang, and  X. F. L\"{u}, Commun. Theor. Phys.
{\bf 37} (2002) 589.
\bibitem{CP75} J. C. Collins, and M. J. Perry, Phys. Rev. Lett. {\bf
34} (1975) 1353.
\bibitem{HK78} P. Hasenfratz, J. Kuti, Phys. Rep. {\bf 40} (1978) 75.
\bibitem{MK01} P. Miskinis, and G. Karlikauskas, Nucl. Phys. {\bf A
683} (2001) 339.
\bibitem{Pra97} M. Prakash, I. Bombaci, M. Prakash, P. J. Ellis,
J. M. Lattimer, and R. Knorren, Phys. Rep. {\bf 280} (1997) 1.

\end{thebibliography}
\end{document}